\documentclass[12pt, a4paper, oneside]{article}

\usepackage[T1]{fontenc}
\usepackage[utf8]{inputenc}
\usepackage{amsmath}
\usepackage{amsfonts}
\usepackage{amsthm}
\usepackage{tabularx}

\usepackage{graphicx}
\usepackage{natbib}

\newcommand{\diffd}{\text{d}}
\newcommand{\fulldiff}[2]{\frac{\diffd #1}{\diffd #2}}

\begin{document}
\title{Quantity restrictions and price discounts on Russian oil\protect}
\author{Henrik Wachtmeister, Johan Gars \& Daniel Spiro\thanks{{\footnotesize{}Order of authors randomized. Wachtmeister: Dept. of
Earth Sciences, Uppsala University, Sweden. Gars: Beijer Institute,
The Royal Swedish Academy of Sciences, Stockholm, Sweden. Spiro: Dept.
of Economics, Uppsala University, Sweden. We thank Nataliia Shapalova,
Borys Dodonov and Jacob Nell for useful comments. Wacthmeister ackonowledges
funding from StandUP for Energy. The theoretical framework and empirical
estimates were partly derived by Gars and Spiro as part of work in
FORMAS project 2020‐00371.}}}
\date{November 2022}
\maketitle
\begin{abstract}
\quad{}Following Russia’s invasion of Ukraine, Western countries
have looked for ways to limit Russia’s oil income. This paper considers,
theoretically and quantitatively, two such options: 1) an export-quantity
restriction and 2) a forced discount on Russian oil. We build a parsimonious
quantifiable model of the global oil market and analyze how each of
these policies affect: which Russian oil fields fall out of production;
the global oil supply; and the global oil price. By these statics
we derive the effects of the policies on Russian oil profits and oil-importers'
economic surplus. The potential effects on Russian oil profits are
substantial. In the short run (within the first year), a quantity
restriction of 20\% yields Russian losses of 62 million USD per day,
equivalent to 1.2\% of GDP and 32\% of military spending. In the long
run (beyond a year) new investments become unprofitable and losses
rise to 100 million USD/day, 2\% of GDP and 56\% of military spending.
A price discount of 20\% is even more harmful for Russia, yielding
losses of 152 million USD per day, equivalent to 3.1\% of GDP and
85\% of military spending in the short run. In the long run, results
are largely the same in terms of Russian profit losses, with the addition
that investment in the oil sector will be harmed. A price discount
puts generally more burden on Russia and less on importers compared
to a quantity restriction. In fact, a price discount implies net gains
for oil importers as it essentially redistributes oil rents from Russia
to importers. If the restrictions are expected to last for long, the
burden on oil importers decreases. Overall, both policies at all levels
imply larger relative losses for Russia than for oil importers (in
shares of their GDP). The case for a price discount on Russian oil
is thus strong, should oil importers want to escalate the economic
warfare with Russia. However, Russia may choose not to export at the
discounted price, in which case the price-discount sanction becomes
a de facto supply restriction.


\end{abstract}

\section{Introduction}

Following Russia's invasion of Ukraine, Western countries have implemented
a large number of economic sanctions on Russia (see \cite{EC_2022}
and \cite{treasury_2022} for lists). In particular, much political
thinking and diplomatic effort has been devoted to tightening sanctions
on the Russian oil sector. This is natural given that total oil revenues
stand for 13\% of (pre-war) Russian GDP, 42\% of Russian exports and
35\% of government income \cite{rystad_energy_ucube_2022,world_bank_exports_2022}.
Yet, since the world economy is reliant on oil, the actual implementation
has remained elusive. Western governments have feared the economic
reprecussions of oil sanctions. As it currently stands, the US and
UK have completely halted their Russian oil imports and the EU has
reduced theirs by 40\% \citep{rystad_energy_specialreport_Russia_2022}
and is planning to implement a near-complete import embargo in the
near term (\cite{weizent_2022}).\footnote{The consequences of the EU import embargo for Russian oil profits
are analyzed by \citet{spiro_import} and the consequences for
the EU by \citep{bachmann2022if} and \citet{berger2022potential}.
See also \citet{gars2022effect} for analysis of the effects on
Russia of other EU energy policies.} In parallel, the G7 countries are discussing a price discount on
Russian oil \citep{Reuters_2022}.\footnote{The G7 proposal is often described as a ``price cap''. According
to the current proposal, Russian oil is supposed to be sold at a fixed
price lower than the current world oil price. More precisely, Russia
will not be able to use Western-controlled tankers and insurance for
transport unless that oil is sold at the discounted price. See \citet{kennedy_2022}
for discussion and \cite{spiro_import} for analysis of a tanker embargo.
In our paper we explore a fixed discount relative to the world oil
price.} Another option to limit Russian oil income is to restrict the quantity
of oil it can export.\footnote{This can be done in a number of direct and indirect ways, e.g., a
limit imports, access to transport and maintenance technology or outright
destruction of infrastructure.}

This paper asks three questions:
\begin{enumerate}
\item What is the effect of a quantity restriction on Russian oil profits
and oil-importers' oil expenditures?
\item What is the effect of a price discount on Russian oil profits and
oil-importers' oil expenditures?
\item How do the burdens on Russia and oil importers compare between the
two policies?
\end{enumerate}
To answer these questions, we build a parsimonious model of the oil
market (adapted from \cite{gars2022effect}, see also \cite{faehn2017climate}
and \cite{erickson2014impact} for similar modeling in other domains).
Both policies imply that some Russian oil fields fall out of production.
This in turn means global oil supply falls which induces a global
oil-price increase and a potential supply increase of other oil producers
as well as a demand respone from consumers. To quantify these effects,
we use oil-field data for Russia and estimates of global demand and
supply quantities and elasticities of oil.

This is thus the first paper to provide a quantitative cost-benefit
analysis of energy sanctions of Russia.\footnote{Previous studies we are aware of study either the cost to Russia \citep{hosoi2022implement,gars2022effect,spiro_import}
or the sanctioning party \citep{bachmann2022if,berger2022potential,lafrogne2022beyond}.
For other conflicts see, e.g., \citet{gharehgozli2017estimation,allen2008domestic,fischhendler2017political,chen2019international,shapovalova2020russian}.} We largely follow the concepts outlined in \citet{johnson1950optimum,sturm2022simple}.

We want to be explicit about an important limitation of this study.
We abstract from the political and diplomatic feasibiliy of the sanctions
and of the precise construction -- we take the restrictions as given
and assume they are complete and enforced. This implies that we abstract
from strategic considerations on the part of Russia and other oil
exporters in their supply of oil. We discuss the strategic implications
in light of our results in the conclusions.

\section{Theoretical framework and results}

Global oil demand is denoted $D(p)$, the Russian oil supply $S_{RU}(p)$
and the oil supply from the rest of the world $S_{ROW}(p)$. All these
are functions of the oil price $p.$ Russian oil supply is given by
a marginal cost curve so that $S_{RU}=MC_{RU}^{-1}$ is the inverse
marginal cost when total extraction is $Q$. We assume a constant
elasticity of supply $\epsilon_{S,ROW}$ for the rest of the world
supply and a constant demand elasticity $\epsilon_{D}$ for total
demand.\footnote{Being interested in sanctions on Russia we thus represent its supply
in more detail while global demand and remaining supply take the simpler
form of constant elasticities.}

Russian oil profits are given by 
\[
\pi_{RU}=pS_{RU}-\int_{0}^{S_{RU}}MC_{RU}(q)\diffd q.
\]

Initially, the equilibrium on the global oil market is determined
by 
\[
D(p)=S_{ROW}(p)+S_{RU}(p)
\]
and the resulting equilibrium price is denoted by $p^{*}$, and the
quantities by $Q^{*}$, $S_{RU}^{*}$ and $S_{ROW}^{*}$. Furthermore,
the share of the oil supply coming from Russia is denoted by $y$
so that 
\[
S_{RU}^{*}=yQ^{*}\text{ and }S_{ROW}^{*}=(1-y)Q^{*}.
\]

In this equilibrium, Russian profits are given by 
\[
\pi_{RU}^{*}=p^{*}S_{RU}^{*}-\int_{0}^{S_{RU}^{*}}MC_{RU}(q)\diffd q.
\]

We are also interested in the consumer surplus associated with oil
use. With a constant elasticity of demand $\epsilon_{D}$, demand
can be written $D(p)=B_{D}p^{\epsilon_{D}}$ where $B_{D}$ is a constant.
The consumer surplus is given by 
\[
CS(p)=\int_{p}^{\infty}D(\rho)\diffd\rho.
\]
With the assumed demand function, the change in consumer surplus when
the oil price changes from $p_{0}$ to $p_{1}$ is 
\begin{equation}
\Delta CS=-\int_{p_{0}}^{p_{1}}B_{D}p^{\epsilon_{D}}\diffd p=-\frac{B_{D}}{\epsilon_{D}+1}\left[p_{1}^{\epsilon_{D}+1}-p_{0}^{\epsilon_{D}+1}\right].\label{Eqn:ConsumerSurplusChange}
\end{equation}

\subsection{Quantity restriction}

We now consider a sanction that reduces Russian oil supply by a share
$\alpha$ compared to their free-trade supply. That is, Russian oil
supply after the sanction has been implemented is 
\[
S_{RU}=(1-\alpha)S_{RU}^{*}.
\]
The rest of the world supply and the demand respond endogenously to
this change.

After the sanction has been implemented, the equilibrium in the oil
market is 
\[
D(p)=S_{ROW}(p)+(1-\alpha)S_{RU}^{*}.
\]
Treating $p$ as a function of $\alpha$, differentiating fully with
respect to $\alpha$ and rewriting gives
\[
\frac{1}{p}\fulldiff{p}\alpha=\frac{y}{(1-y)\epsilon_{S,ROW}-\epsilon_{D}}.
\]
The price change following a sanction of $\alpha$ can be approximated
by 
\[
\Delta_{\alpha}p\approx p^{*}\left(e^{\xi\alpha}-1\right),
\]

where

\[
\xi\equiv\frac{y}{(1-y)\epsilon_{S,ROW}-\epsilon_{D}}
\]

is a constant.

The resulting change in Russian oil profits can then be approximated
by

\begin{equation}
\Delta_{\alpha}\pi_{RU}\approx\left(\left(1-\alpha\right)\Delta_{\alpha}p-\alpha p^{*}\right)S_{RU}^{*}+\int_{(1-\alpha)S_{RU}^{*}}^{S_{RU}^{*}}MC_{RU}(q)\diffd q.\label{Eqn:ProfitLossQuantity}
\end{equation}
Since the sanction will lead to an increased oil price, $\Delta_{\alpha}p>0$,
Russia will gain from an increase in the oil price and from a decrease
of extraction costs but lose from the decreased quantity. For relatively
small $\alpha$, the effect on extraction costs will be relatively
large since the marginal oil is expensive to extract. Hence the net
effect could theoretically be positive for small $\alpha$. Whether
they are so in practice will be examined quantitatively.

Using (\ref{Eqn:ConsumerSurplusChange}) the change in consumer surplus
is

\begin{equation}
\Delta_{\alpha}CS\approx-\int_{p^{*}}^{p^{*}+\Delta_{\alpha}p}D(p)=-\frac{B}{\epsilon_{D}+1}\left[\left(p^{*}+\Delta_{\alpha}p\right)^{\epsilon_{D}+1}-\left(p^{*}\right)^{\epsilon_{D}+1}\right].\label{Eqn: CS quantity}
\end{equation}

\subsection{Price discount}

We now consider a price sanction that allows Russia to sell as much
oil as it wants but only at a discount $\delta$ compared to the current
global oil price. With this sanction, the equilibrium in the oil market
is
\[
D(p)=S_{ROW}(p)+S_{RU}((1-\delta)p).
\]
When approximating the effect of a sanction $\delta$ we assume that
demand and supply from ROW can be approximated linearly based on elasticities
while the Russian oil supply cannot since the relative change is bigger
there and the exact shape of the supply curve matters more. The resulting
price change $\Delta_{\delta}p$ is then implicitly given by the approximation
\[
\epsilon_{D}\Delta_{\delta}p\approx(1-y)\epsilon_{S,ROW}\Delta_{\delta}p+yp^{*}\frac{S_{RU}\left((1-\delta)(p^{*}+\Delta_{\delta}p)\right)-S_{RU}^{*}}{S_{RU}^{*}}.
\]

Denote the resulting approximate change in Russian oil supply by 
\[
\Delta_{\delta}S_{RU}\equiv S_{RU}\left((1-\delta)(p^{*}+\Delta_{\delta}p)\right)-S_{RU}^{*}.
\]

We can now approximate the difference in Russian oil profits due to
a price discount by 
\begin{equation}
\Delta_{\delta}\pi_{RU}\approx\left((1-\delta)\Delta_{\delta}p-\delta p^{*}\right)S_{RU}^{*}+(1-\delta)\left(p^{*}+\Delta_{\delta}p\right)\Delta_{\delta}S_{RU}+\int_{S_{RU}^{*}+\Delta_{\delta}S_{RU}}^{S_{RU}^{*}}MC_{RU}(q)\diffd q.\label{Eqn:ProfitLossPrice}
\end{equation}

When computing the change in consumer surplus there are now two effects.
The first is a change in the market price $p$. This part is the same
as in (\ref{Eqn:ConsumerSurplusChange}). The second effect is that
the oil bought from Russia is bought at a discount $\delta$. This
second effect gives an increase in consumer surplus. The joint effect
is 
\begin{equation}
\Delta_{\delta}CS=-\frac{B_{D}}{\epsilon_{D}+1}\left[\left(p^{*}+\Delta_{\delta}p\right)^{\epsilon_{D}+1}-\left(p^{*}\right)^{\epsilon_{D}+1}\right]+\delta\left(S_{RU}^{*}+\Delta_{\delta}S_{RU}\right)\left(p^{*}+\Delta_{\delta}p\right)\label{Eqn: CS price}
\end{equation}
that is, we assume all buyers of oil get their ``fair'' share of
discounted oil.

\section{Model parameters and quantities}

The parameters needed to quantitatively assess the equilibrium are
summarized in Table \ref{Table: Parameters and quantities }. Several
are the same as used in \citet{gars2022effect} and are motivated
there.

\begin{table}
\rule[0.5ex]{1\columnwidth}{1pt}

\caption{\label{Table: Parameters and quantities }Parameters and quantities}

\rule[0.5ex]{1\columnwidth}{1pt}
\centering{}{\footnotesize{}}%
\begin{tabular}{lcccc}
\hline 
{\scriptsize{}Parameter} &  & {\scriptsize{}Value Short run} & {\scriptsize{}Value Long run} & {\scriptsize{}Reference}\tabularnewline
\hline 
{\scriptsize{}World oil demand} & {\scriptsize{}$D(p)$} & {\scriptsize{}99 Mb/d} & {\scriptsize{}99 Mb/d} & {\scriptsize{}Rystad UCube database \citep{rystad_energy_ucube_2022}}\tabularnewline
{\scriptsize{}Oil supply Russia (exports)} & {\scriptsize{}$S_{RU}(p)$} & {\scriptsize{}7.5 Mb/d} & {\scriptsize{}7.5 Mb/d} & {\scriptsize{}Rystad UCube database \citep{rystad_energy_ucube_2022}}\tabularnewline
{\scriptsize{}Elasticity of supply Russia} & {\scriptsize{}$\epsilon_{S,ROW}$} & {\scriptsize{}See Figure \ref{Fig: Supply function}} & {\scriptsize{}See Figure \ref{Fig: Supply function}} & {\scriptsize{}Rystad UCube database \citep{rystad_energy_ucube_2022}}\tabularnewline
{\scriptsize{}Oil supply ROW (incl. RU domestic)} & {\scriptsize{}$S_{ROW}(p)$} & {\scriptsize{}91.5 Mb/d} & {\scriptsize{}91.5 Mb/d} & {\scriptsize{}Rystad UCube database \citep{rystad_energy_ucube_2022}}\tabularnewline
{\scriptsize{}Elasticity of supply ROW} & {\scriptsize{}$\epsilon_{S,ROW}$} & {\scriptsize{}0} & {\scriptsize{}0.13} & {\scriptsize{}\citet{gars2022effect}}\tabularnewline
{\scriptsize{}Elasticity of demand} & {\scriptsize{}$\epsilon_{D}$} & {\scriptsize{}-0.125} & {\scriptsize{}-0.45} & {\scriptsize{}\citet{gars2022effect}}\tabularnewline
\hline 
\end{tabular}{\footnotesize\par}
\end{table}

\begin{figure}[h]
\caption{\label{Fig: Supply function}Field based supply function, Russia}

\begin{centering}
\includegraphics[width=0.6\textwidth]{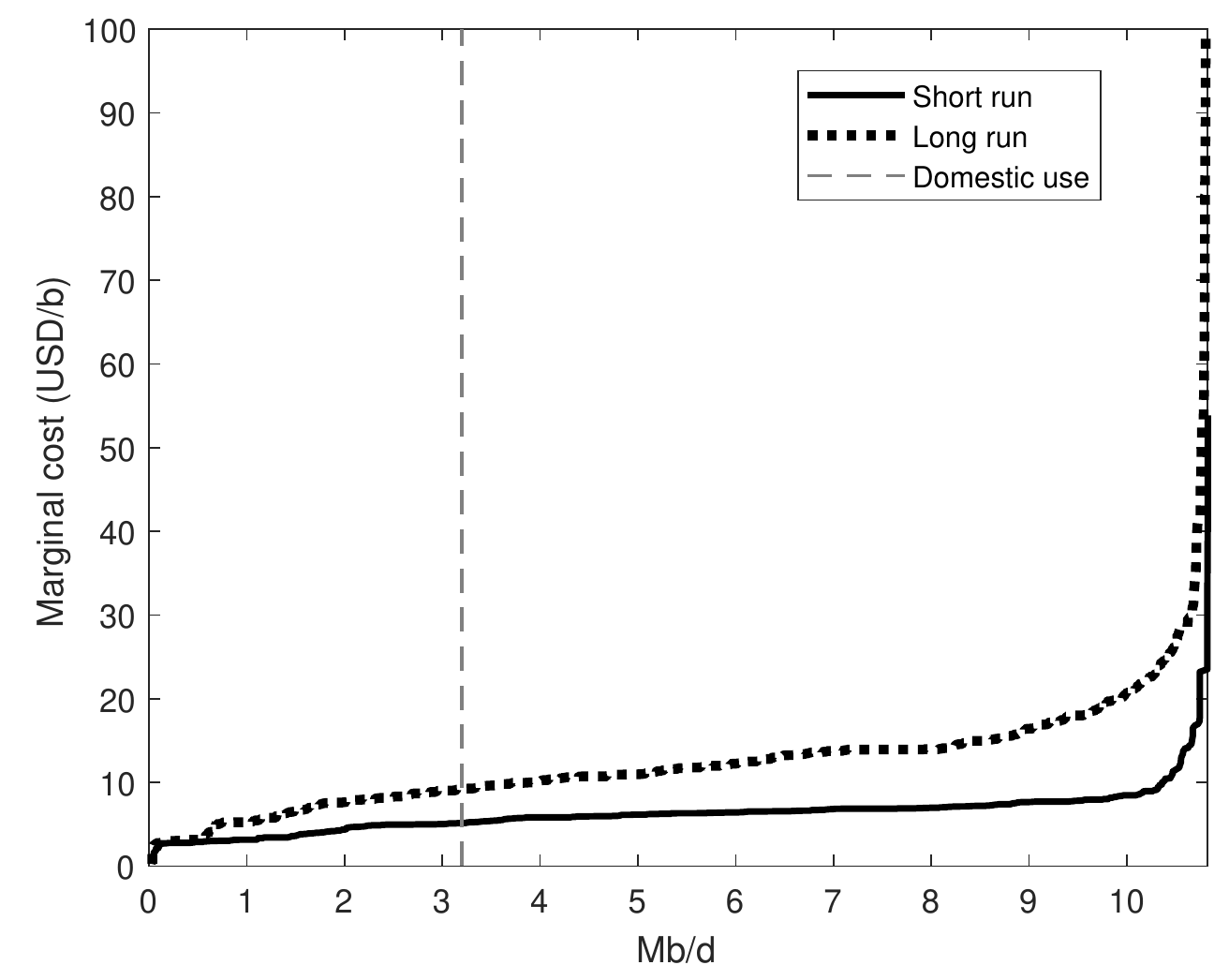}
\par\end{centering}
\raggedright{}{\footnotesize{}Notes: Marginal cost curves based on
field-by-field in Russia from \citet{rystad_energy_ucube_2022}.
The short run curve includes operating costs (OPEX). The long-run
curve includes also investment costs (CAPEX). Government take is not
included.}{\footnotesize\par}
\end{figure}

An important exception is the supply function of Russia. For this
we use data on field-by-field costs from Rystad Energy Ucube database
\citep{rystad_energy_ucube_2022}. These supply functions, one
for the short run and one for the long run, are depicted in Figure
\ref{Fig: Supply function}.

We distinguish in our analysis between short and long run effects
of the policies. Short run can be thought of as within a first year
and long run as beyond one year and up to three years.
The difference is captured in the parameters, where elasticities generally
are higher in the long run (see Table \ref{Table: Parameters and quantities }).
Importantly, we differentiate between the Russian supply functions
in the short and long run. In the short run, investment costs (CAPEX)
that have gone into finding and developing a field are sunk. What
remains are only the costs of operation (OPEX). In the long run new
fields need to be found and developed and the choice of this depends
on the expected price. Thus, in the long run CAPEX are \textit{not}
sunk but part of lowering the profit margins. The conceptual difference
between the short run and long run is therefore that in the short
run, the restrictions imposed happen as a surprise while in the long
run the restrictions are part of affecting expectations.\footnote{In practice, the shift between short and long run is smooth and represented
by a gradual increase in new fields as share of all fields, thus a
gradual increase in the fields for which CAPEX are part of determining
profits.} None of the supply curves include Russian government take since we
are interested in the effects on the Russian economy as a whole.

\section{Results}

\subsection{Short run}

\begin{figure}[h]
\caption{\label{Fig: Q short run}Quantity restriction, Short run}

\begin{centering}
\includegraphics[height=0.5\textheight]{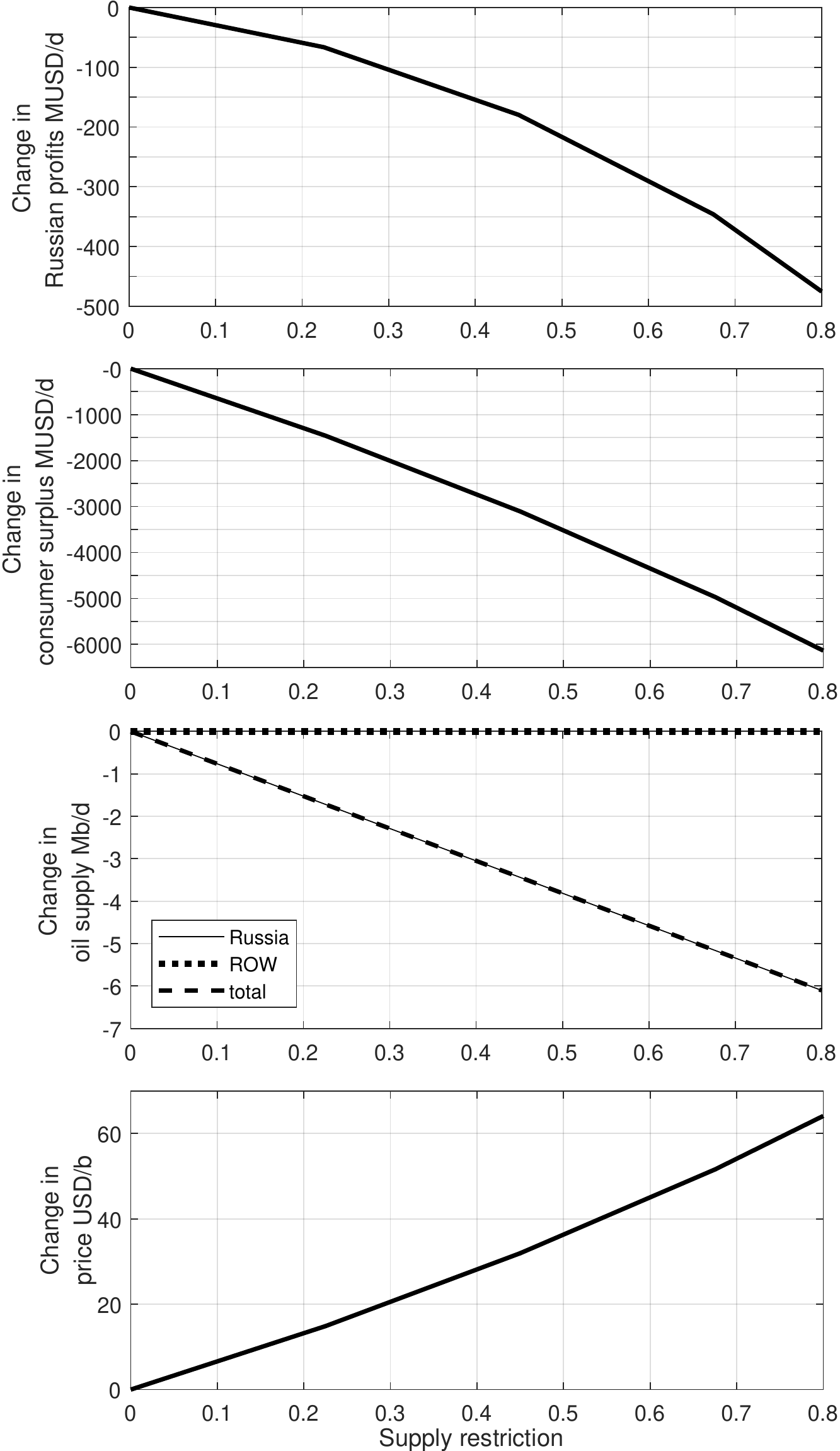}
\par\end{centering}
\raggedright{}{\footnotesize{}Notes: Effects of a quantity restriction
in the short run. x-axis in all panels is extent of restriction. First
panel: Change in Russian profit. Second panel: Change of oil-consumer
surplus. Third panel: Change in quantity from Russia, from other oil
exporters and in aggregate. Fourth panel: Change of oil price.}{\footnotesize\par}
\end{figure}

The effects of a \textbf{quantity restriction in the short run} are
depicted in Figure \ref{Fig: Q short run}. The upper panel shows
Russia's profit losses (from (\ref{Eqn:ProfitLossQuantity})) and
the second panel the oil-consumer' surplus (from (\ref{Eqn: CS quantity})).
We will interpret the latter as a proxy for oil-importers' gains and
losses from the policy and will use consumer surplus and oil-importers'
surplus interchangeably. As can be seen, both are falling as the quantity
restriction becomes tighter. This is since there is a fall in aggregate
supply driven by the reduced Russian export (third panel) which leads
to a substantial price increase (fourth panel). The supply and price
reactions are large since, in the short run, supply from the remaining
producers is inelastic. A quantity restriction of 20\% yields Russian
profit losses of 57 million USD/day, a 30\% restriction yields 99
million USD/day and 50\% yields 212 million USD/day.

\begin{figure}[h]
\caption{\label{Fig: P short run}Price discount, Short run}

\begin{centering}
\includegraphics[height=0.5\textheight]{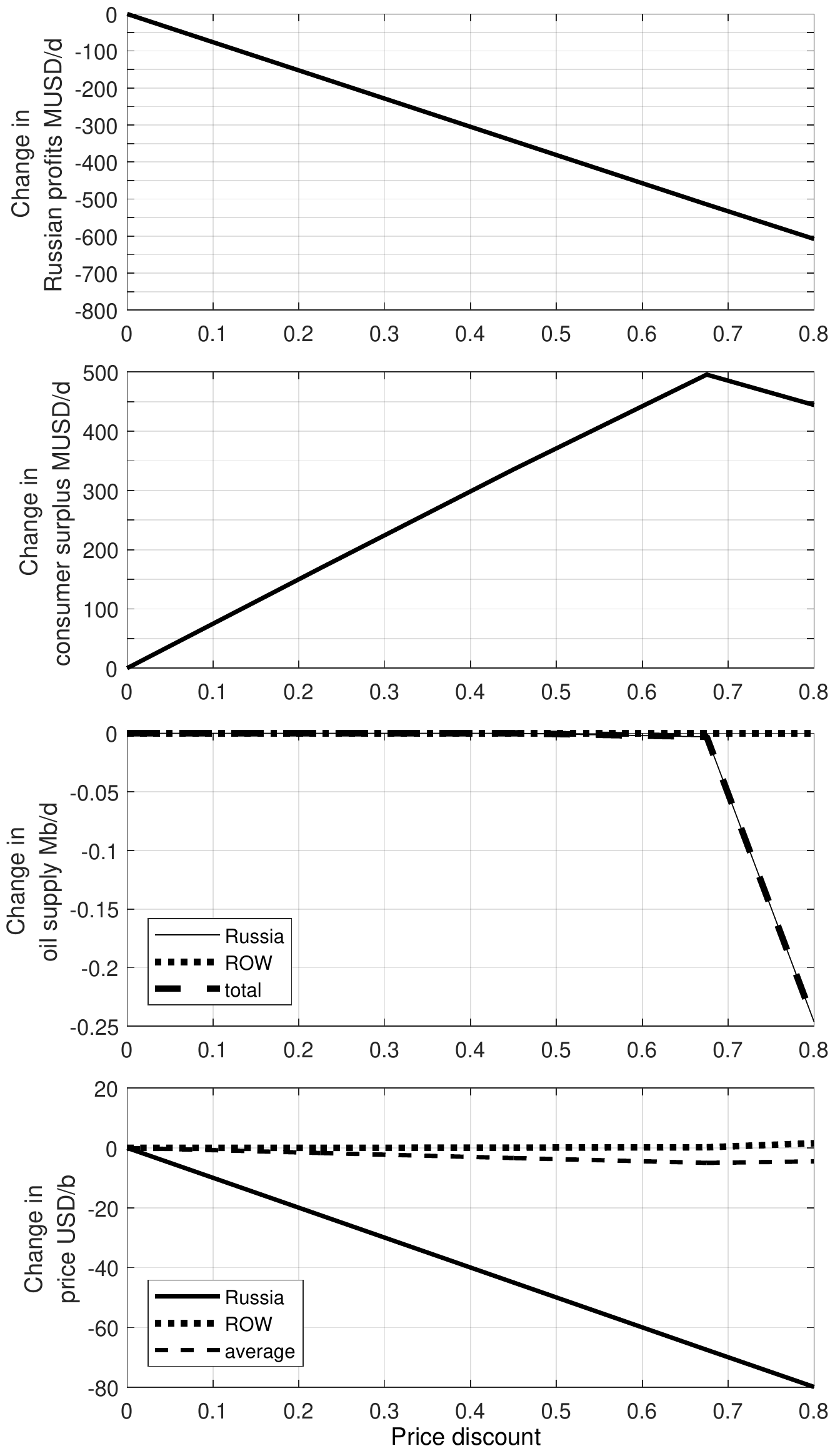}
\par\end{centering}
\raggedright{}{\footnotesize{}Notes: Effects of a price discount in
the short run. x-axis in all panels is extent of restriction. First
panel: Change in Russian profit. Second panel: Change of oil-consumer
surplus. Third panel: Change in quantity from Russia, from other oil
exporters and in aggregate. Fourth panel: Change of oil price Russia
gets, global oil price and weighted average.}{\footnotesize\par}
\end{figure}

The effects of a \textbf{price discount in the short run} are depicted
in Figure \ref{Fig: P short run}. The upper panel shows Russia's
profit losses (from (\ref{Eqn:ProfitLossPrice})) and the second panel
the change in consumer surplus (from (\ref{Eqn: CS price})). As can
be seen, a price discount leads to Russian losses and an \textit{increase}
in consumer surplus. This is since, in effect, the price discount
is equivalent to a succesful buyers' cartel. An important detail is
that, under the price discount, Russia does not reduce its supplies
until the discount is above roughly 75\% (third panel). This is since
Russian oil is largely profitable even at very low oil prices, in
the short run. This means that there is no global oil-price reaction
until that level either (fourth panel). At around 75\% discount, Russian
production starts falling out. Since other oil producers are inelastic
in the short run, this spurs an increase in the global oil price.
This is what makes consumer surplus start falling at around 80\% discount.
Nevertheless Russian profits keep decreasing as the discount is strong
enough to offset that the increased global oil price may drive up
the Russian (discounted) price. A price discount of 20\% yields Russian
profit losses of 152 million USD/day, a 30\% restriction yields 229
million USD/day and 50\% yields 381 million USD/day.

\subsection{Long run}

\begin{figure}[h]
\caption{\label{Fig: Q long run}Quantity restriction, Long run}

\begin{centering}
\includegraphics[height=0.5\textheight]{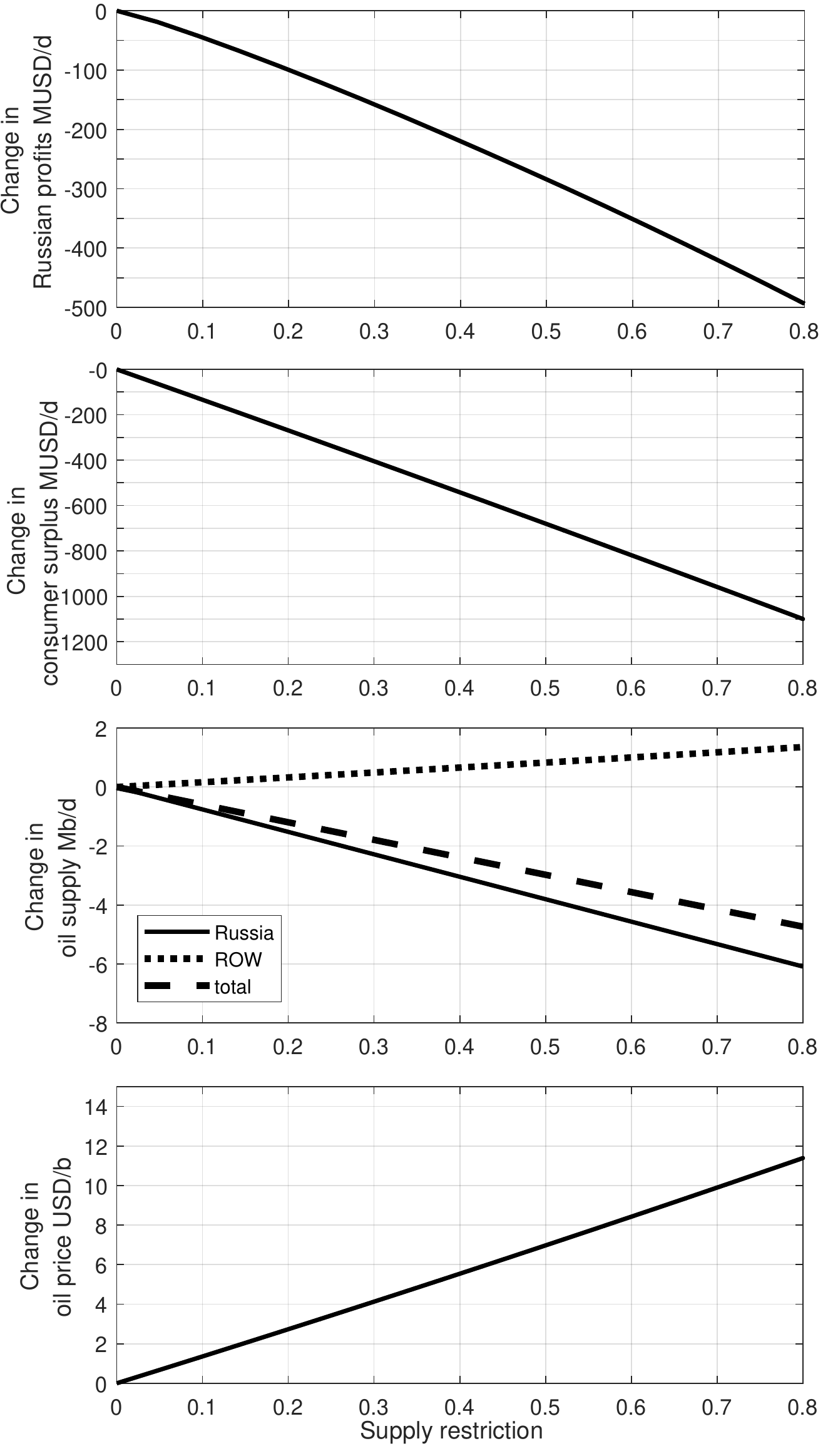}
\par\end{centering}
\raggedright{}{\footnotesize{}Notes: Effects of a quantity restriction
in the long run. x-axis in all panels is extent of restriction. First
panel: Change in Russian profit . Second panel: Change of oil-consumer
surplus. Third panel: Change in quantity from Russia, from other oil
exporters and in aggregate. Fourth panel: Change of oil price.}{\footnotesize\par}
\end{figure}

The effects of a \textbf{quantity restriction in the long run} are
depicted in Figure \ref{Fig: Q long run}. Russian profits fall monotonically
with the extent of the restriction and so does the consumer surplus.
Qualitatively it is thus similar to what happens in the short run.
Quantitatively there is a large difference though. Russian losses
are larger: a quantity restriction of 20\% in the long run yields
Russian profit losses of 100 million USD/day, a 30\% restriction yields
158 million USD/day and 50\% yields 284 million USD/day. In the long
run the oil importers' surplus reduction is substantially smaller.
The reason for this is partly since the elasticity of other suppliers
is much higher, meaning that they substitute some of the Russian oil
that falls out. This is visible in the third panel where supply from
other producers is increasing. Partly, the lower burden on oil importers
is also due to demand being more elastic, implying that buyers substitute
away from oil. For this reason the relative burden of quantity restrictions
is higher on Russia in the long run.

\begin{figure}[h]
\caption{\label{Fig: P long run}Price discount, Long run}

\begin{centering}
\includegraphics[height=0.5\textheight]{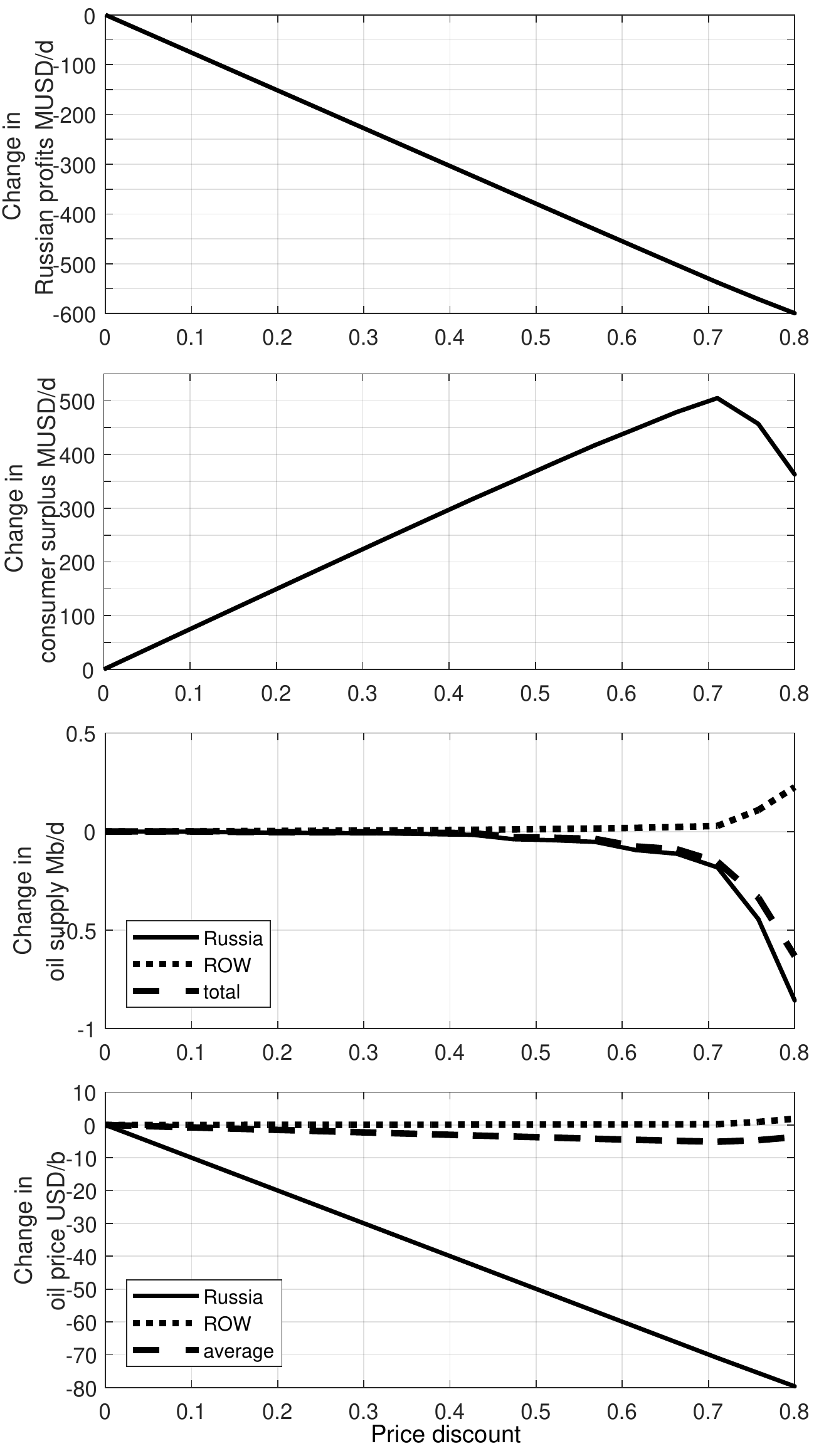}
\par\end{centering}
\raggedright{}{\footnotesize{}Notes: Effects of a price discount in
the long run. x-axis in all panels is extent of restriction. First
panel: Change in Russian profit . Second panel: Change of oil-consumer
surplus. Third panel: Change in quantity from Russia, from other oil
exporters and in aggregate. Fourth panel: Change of oil price Russia
gets, global oil price and weighted average.}{\footnotesize\par}
\end{figure}

The effects of a \textbf{price discount in the long run }are depicted
in Figure \ref{Fig: P long run}. Russian profits are falling monotonically
with the discount. A price discount of 20\% yields Russian profit
losses of 152 million USD/day, a 30\% restriction yields 228 million
USD/day and 50\% yields 379 million USD/day. In the long run, Russian
supply falls already at a discount of around 45\%. This is earlier
than in the short run since new field investments are not profitable
for a longer-lasting discount. With the more elastic supply from remaining
producers, total supply falls less than in the short run. Nevertheless
the reduced supply at smaller discounts implies that the consumer
surplus reaches its peak earlier.

\subsection{Who bears the burdens of the policies?}

\begin{figure}[h]
\caption{\label{Fig: P vs Q Short run}Comparison of Quantity restriction and
Price discount}

\begin{centering}
\includegraphics[clip,height=15cm]{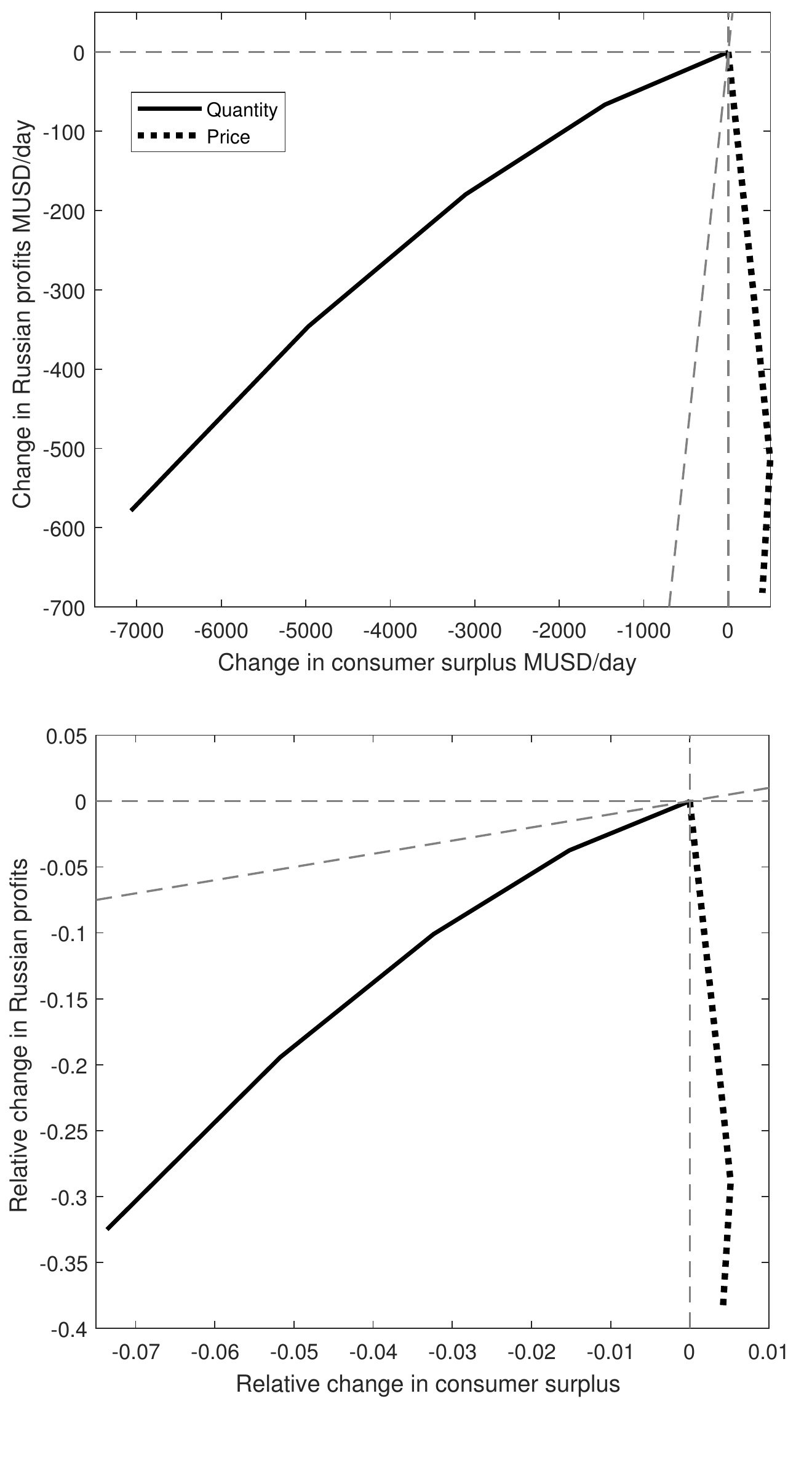}\includegraphics[clip,height=15cm]{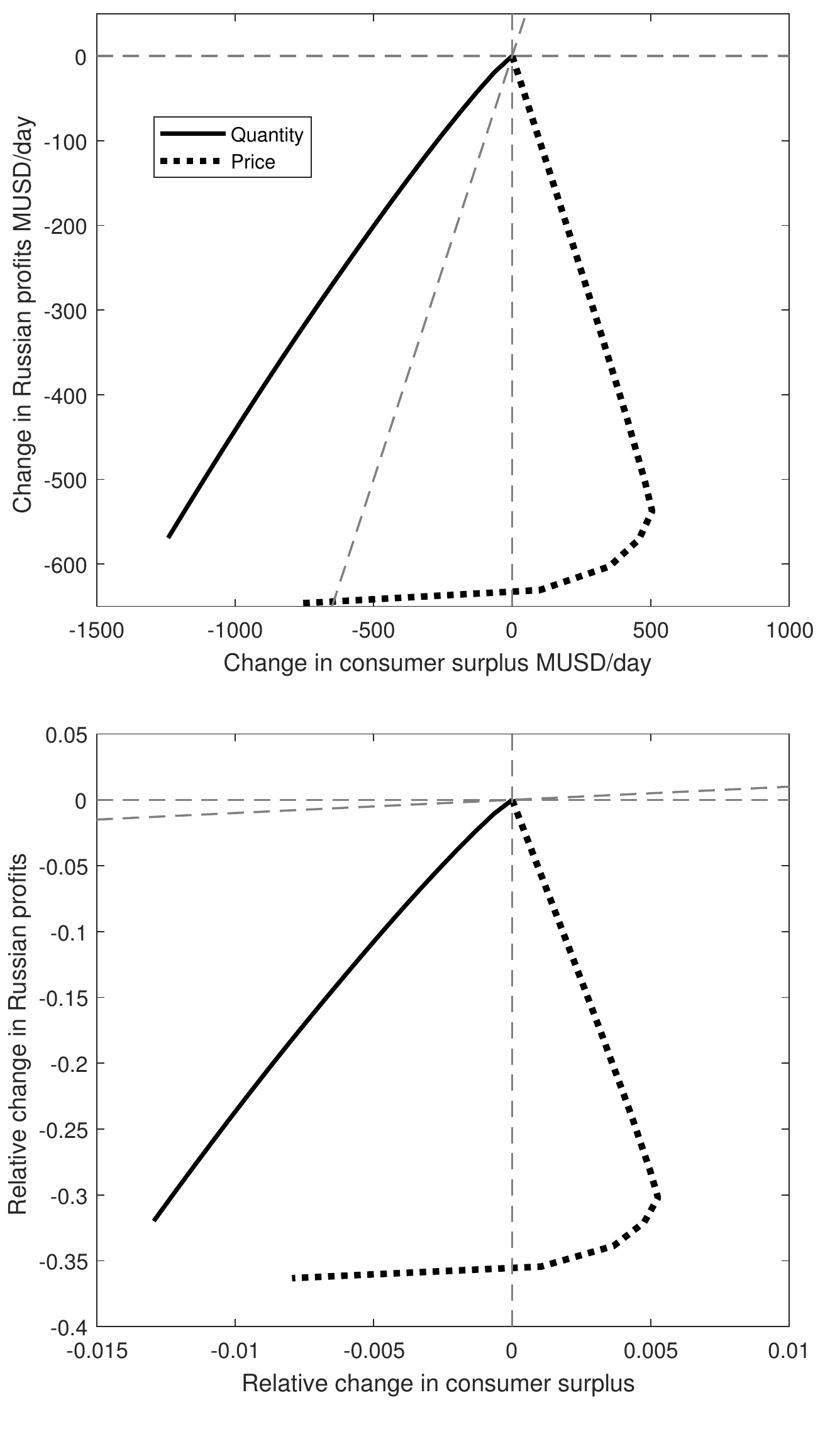}
\par\end{centering}
\raggedright{}{\footnotesize{}Notes: Comparison of policies in the
short run (left panels) and long run (right panels). Change in oil-consumers'
surplus (x-axis) and Russian profits (y-axis) for the two policies.
Upper panel is in USD and lower panel is USD/GDP or each respectively.
The vertical and horizontal lines mark the zero-lines, the sloping
dashed line is the 45-degree-line marking equal burden.}{\footnotesize\par}
\end{figure}

Figure \ref{Fig: P vs Q Short run} \textbf{compares the two policies}.
The horizontal axes represent consumer surplus and the vertical axes
show the Russian loss. The upper-panel axes are measured in USD while
the lower-panel axes are measured in shares of GDP (Russia's pre-war
GDP on the y-axis and global GDP on the x-axis). Gains and losses
normalized by GDP provide important information since, in war, the
adversaries' relative losses matter. The left panels are in the short
run while the right panels are in the long run. We can note that the
x-axis covers both positive values -- representing gains for oil
importers -- and negative values. The y-axis has only negative values
since Russia loses from both policies. Furthermore, for the price-discount
policy there may exist two values of losses for Russia for the same
consumer gain. This is since the effect of policy on consumer surplus
is non-monotonic.

What is evident from the figure is that the price discount generally
puts more burden on Russia compared to the quantity restriction. For
any given Russian loss (fixing a point on the y-axis), importers gain
more with the price discount than with the quantity restriction.

Furthermore, in absolute terms, the supply restriction is north-west
of the 45-degree line implying that oil importers bear a heavier burden.
The price discount, on the other hand, is mostly to the south-east
of the 45-degree line implying that Russia bears a heavier burden,
except for a very extensive price discount which is captured by the
part of the line that is to the right of the 45-degree line. Taking
into account the economic size of Russia viz-a-viz the oil importers,
Russia bears a much heavier burden than oil importers for both supply
restrictions and the price discount.

\subsection{Regional breakdown of consumer surplus}

A contentious issue before and during the war has been the effect
of various sanctions on the population, i.e., consumers in the sanctioning
countries. We here report the change in consumer surplus in four main
economies: US, EU, India and China. The two former have been part
of sanctions and the two latter have not as of yet.

Table \ref{Table: Regions quantity} reports, for quantity restrictions,
the change of consumer surplus as share of these regions GDP. The
results are based on equation (\ref{Eqn: CS quantity}), and take
into account the oil consumption ($B_{D}$) of the economies.

The table indicates that, of the potential sanctioning economies,
India bears the largest relative burden. This is since its economy
is more oil intensive than the others'. The consumer loss is more
pronounced in the short run, indicating that the main political obstacle
is getting the sanctioning in place rather than making it last in
the long term. 

The table also shows that Russia's economy will be harmed relatively
more for any level of sanctioning. In particular in the long run,
where their losses are an order magnitude larger than the oil consumers'. 

\begin{table}
\rule[0.5ex]{1\columnwidth}{1pt}

\caption{\label{Table: Regions quantity}Change in consumer surplus across
regions -{}- quantity restriction}

\rule[0.5ex]{1\columnwidth}{1pt}
\begin{centering}
{\footnotesize{}}%
\begin{tabular}{l}
\hline 
\multicolumn{1}{c}{{\scriptsize{}Extent ($\alpha$)}}\tabularnewline
\hline 
\tabularnewline
\hline 
{\scriptsize{}10\%}\tabularnewline
{\scriptsize{}30\%}\tabularnewline
{\scriptsize{}50\%}\tabularnewline
{\scriptsize{}70\%}\tabularnewline
\hline 
\end{tabular}{\footnotesize{}}%
\begin{tabular}{lccccc}
\hline 
\multicolumn{6}{c}{{\scriptsize{}Short run}}\tabularnewline
\hline 
 & {\scriptsize{}EU} & {\scriptsize{}US} & {\scriptsize{}India} & {\scriptsize{}China} & {\scriptsize{}Russia}\tabularnewline
\hline 
 & {\scriptsize{}-0.15\%} & {\scriptsize{}-0.20\%} & {\scriptsize{}-0.36\%} & {\scriptsize{}-0.21\%} & {\scriptsize{}-0.47\%}\tabularnewline
 & {\scriptsize{}-0.47\%} & {\scriptsize{}-0.63\%} & {\scriptsize{}-1.1\%} & {\scriptsize{}-0.65\%} & {\scriptsize{}-2.0\%}\tabularnewline
 & {\scriptsize{}-0.82\%} & {\scriptsize{}-1.1\%} & {\scriptsize{}-2.0\%} & {\scriptsize{}-1.2\%} & {\scriptsize{}-4.3\%}\tabularnewline
 & {\scriptsize{}-1.1\%} & {\scriptsize{}-1.6\%} & {\scriptsize{}-3.0\%} & {\scriptsize{}-1.7\%} & {\scriptsize{}-7.6\%}\tabularnewline
\hline 
\end{tabular}{\footnotesize{}}%
\begin{tabular}{lccccc}
\hline 
\multicolumn{6}{c}{{\scriptsize{}Long run}}\tabularnewline
\hline 
 & {\scriptsize{}EU} & {\scriptsize{}US} & {\scriptsize{}India} & {\scriptsize{}China} & {\scriptsize{}Russia}\tabularnewline
\hline 
 & {\scriptsize{}-0.03\%} & {\scriptsize{}-0.04\%} & {\scriptsize{}-0,08\%} & {\scriptsize{}-0.04\%} & {\scriptsize{}-0.9\%}\tabularnewline
 & {\scriptsize{}-0.10\%} & {\scriptsize{}-0.13\%} & {\scriptsize{}-0.23\%} & {\scriptsize{}-0.13\%} & {\scriptsize{}-3.2\%}\tabularnewline
 & {\scriptsize{}-0.16\%} & {\scriptsize{}-0.22\%} & {\scriptsize{}-0.39\%} & {\scriptsize{}-0.22\%} & {\scriptsize{}-5.8\%}\tabularnewline
 & {\scriptsize{}-0.23\%} & {\scriptsize{}-0.30\%} & {\scriptsize{}-0.55\%} & {\scriptsize{}-0.32\%} & {\scriptsize{}-8.7\%}\tabularnewline
\hline 
\end{tabular}{\footnotesize\par}
\par\end{centering}
{\small{}Notes: Change of consumer surplus as share of GDP due to
a quantity restriction. GDP data (USD in 2021) from worldbank.org.
Oil consumption data from \citet{bp_bp_2022}. Left table is short
run, right table is long run. Right colum shows Russia's profit loss
as share of GDP.}{\small\par}
\end{table}

Table \ref{Table: Regions price} reports, for price discounts, the
change of consumer surplus as share of these regions' GDP. The results
are based on equation (\ref{Eqn: CS price}), and take into account
the oil consumption ($B_{D}$) of the economies. The table reflects
that oil consumers gain from a price discount, and more so the more
extensive the discount is. Recall, however, from Figure \ref{Fig: P short run}
that the optimal discount is at around 70-80\%. Consumer surplus decreases
beyond that level.

\begin{table}
\rule[0.5ex]{1\columnwidth}{1pt}

\caption{\label{Table: Regions price}Change in consumer surplus across regions
-{}- price discount}

\rule[0.5ex]{1\columnwidth}{1pt}
\begin{centering}
{\footnotesize{}}%
\begin{tabular}{l}
\hline 
\multicolumn{1}{c}{{\scriptsize{}Extent ($\delta$)}}\tabularnewline
\hline 
\tabularnewline
\hline 
{\scriptsize{}10\%}\tabularnewline
{\scriptsize{}30\%}\tabularnewline
{\scriptsize{}50\%}\tabularnewline
{\scriptsize{}70\%}\tabularnewline
\hline 
\end{tabular}{\footnotesize{}}%
\begin{tabular}{lccccc}
\hline 
\multicolumn{6}{c}{{\scriptsize{}Short run}}\tabularnewline
\hline 
 & {\scriptsize{}EU} & {\scriptsize{}US} & {\scriptsize{}India} & {\scriptsize{}China} & {\scriptsize{}Russia}\tabularnewline
\hline 
 & {\scriptsize{}0.02\%} & {\scriptsize{}0.02\%} & {\scriptsize{}0.04\%} & {\scriptsize{}0.02\%} & {\scriptsize{}-1.6\%}\tabularnewline
 & {\scriptsize{}0.05\%} & {\scriptsize{}0.07\%} & {\scriptsize{}0.13\%} & {\scriptsize{}0.07\%} & {\scriptsize{}-4.7\%}\tabularnewline
 & {\scriptsize{}0.09\%} & {\scriptsize{}0.12\%} & {\scriptsize{}0.21\%} & {\scriptsize{}0.12\%} & {\scriptsize{}-7.8\%}\tabularnewline
 & {\scriptsize{}0.12\%} & {\scriptsize{}0.16\%} & {\scriptsize{}0.3\%} & {\scriptsize{}0.17\%} & {\scriptsize{}-11\%}\tabularnewline
\hline 
\end{tabular}{\footnotesize{}}%
\begin{tabular}{lccccc}
\hline 
\multicolumn{6}{c}{{\scriptsize{}Long run}}\tabularnewline
\hline 
 & {\scriptsize{}EU} & {\scriptsize{}US} & {\scriptsize{}India} & {\scriptsize{}China} & {\scriptsize{}Russia}\tabularnewline
\hline 
 & {\scriptsize{}0.02\%} & {\scriptsize{}0.02\%} & {\scriptsize{}0.04\%} & {\scriptsize{}0.02\%} & {\scriptsize{}-1,6\%}\tabularnewline
 & {\scriptsize{}0.05\%} & {\scriptsize{}0.07\%} & {\scriptsize{}0.13\%} & {\scriptsize{}0.07\%} & {\scriptsize{}-4.7\%}\tabularnewline
 & {\scriptsize{}0.07\%} & {\scriptsize{}0.12\%} & {\scriptsize{}0.21\%} & {\scriptsize{}0.12\%} & {\scriptsize{}-7.8\%}\tabularnewline
 & {\scriptsize{}0.12\%} & {\scriptsize{}0.16\%} & {\scriptsize{}0.29\%} & {\scriptsize{}0.17\%} & {\scriptsize{}-11\%}\tabularnewline
\hline 
\end{tabular}{\footnotesize\par}
\par\end{centering}
{\small{}Notes: Change of consumer surplus as share of GDP due to
a price discount. GDP data (USD in 2021) from worldbank.org. Oil consumption
data from \citet{bp_bp_2022}. Left table is short run, right table
is long run. Right colum shows Russia's profit loss as share of GDP.}{\small\par}
\end{table}

Jointly, Tables \ref{Table: Regions quantity} and \ref{Table: Regions price}
suggest that all regions would prefer a price discount over a quantity
restriction.

\subsection{Which policy would Russia dislike the least?}

\begin{figure}[h]
\caption{\label{Fig: P vs Q Russia}Quantity restriction and Price discount
from Russia's point of view}

\begin{centering}
\includegraphics[width=0.5\textwidth]{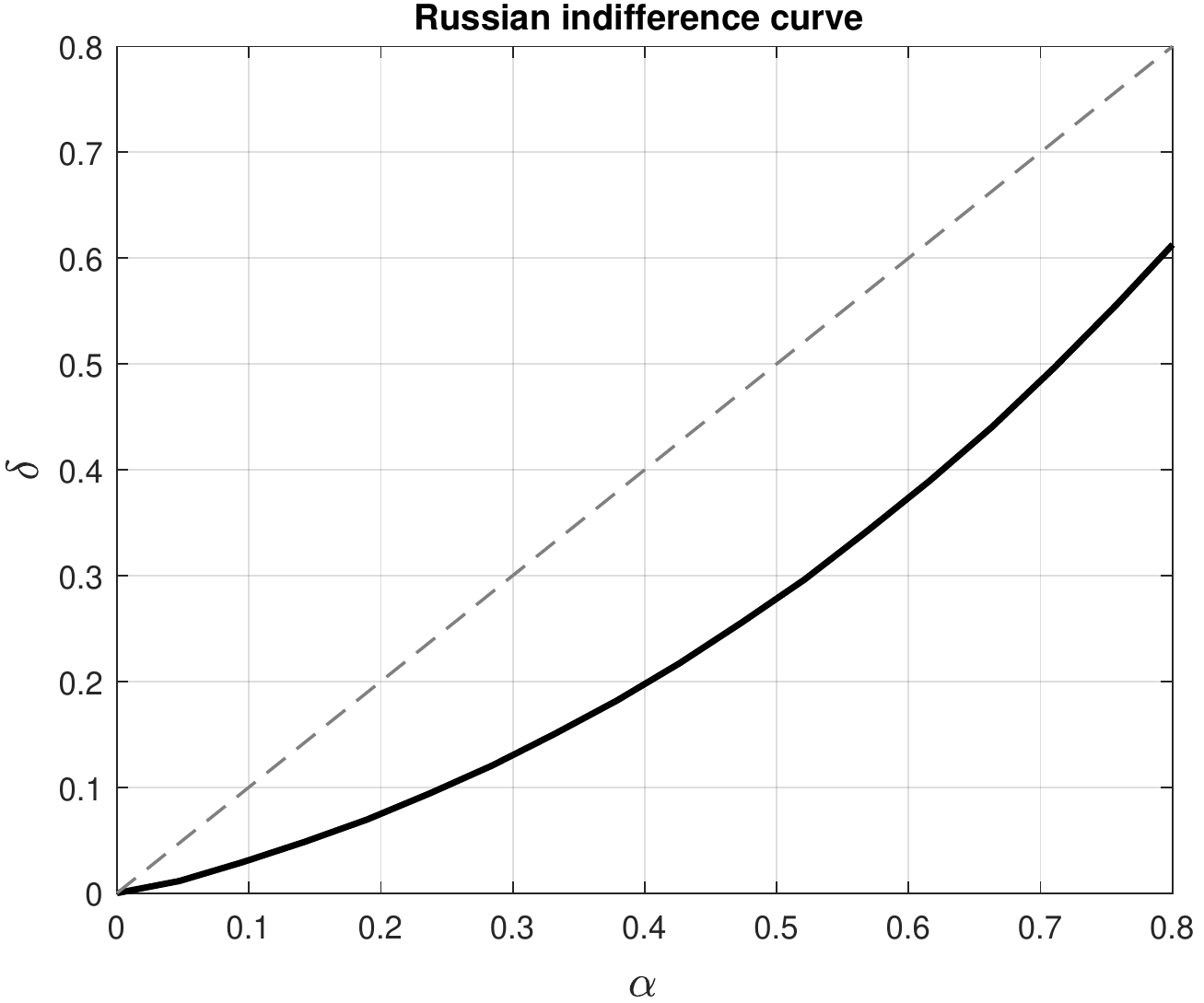}\includegraphics[width=0.5\textwidth]{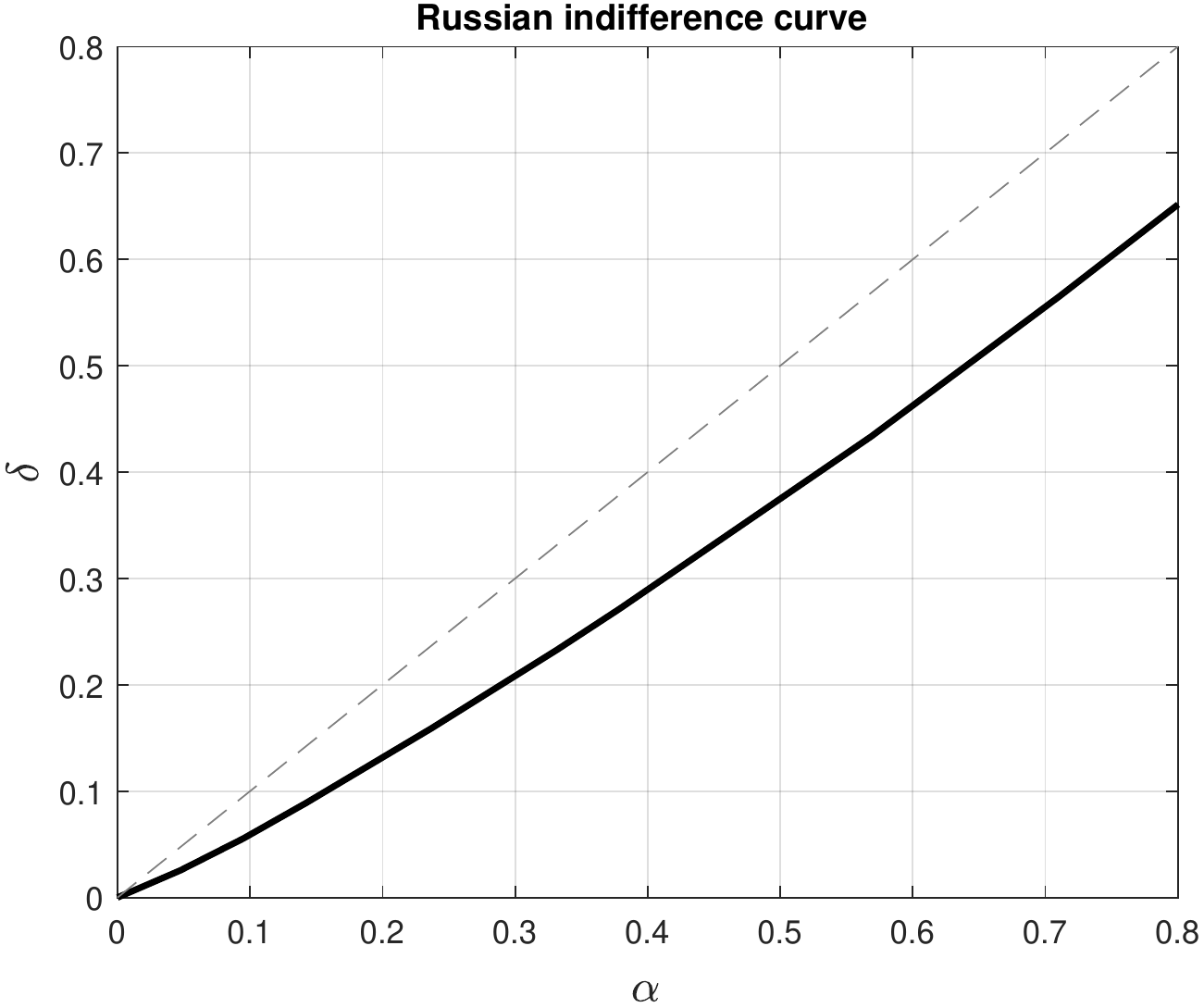}
\par\end{centering}
\raggedright{}{\footnotesize{}Notes: Comparison of Russian losses.
Each curve shows the extent of quantity restriction (x-axis) that
would make Russia indifferent to a price discount at a certain level
(y-axis). Dashed line is 45-degree line. Left is short run and right
is long run.}{\footnotesize\par}
\end{figure}

Figure \ref{Fig: P vs Q Russia} compares the policies from Russia's
point of view. It shows an indifference curve between the policies.
For each percentage of quantity restriction (x-axis) the figure shows
the price discount (y-axis) for which Russia is equally harmed. To
the north-west of the line are combinations of price discount and
quantity restriction such that Russia would prefer the quantity restriction.
As an example, in the short run, if the world imposes a price discount
of 20\%, Russia would rather limit their supply by any share up to
roughly 40\%. Since the curves generally reside below the 45-degree-line,
it means that a price discount is generally more harmful to Russia
than the same level of quantity restriction.

\section{\label{sec:Conclusion}Conclusions and discussion}

We find that both a supply restriction and price discount on Russian
oil imply substantial economic losses for Russia. For example, a supply
restriction covering 20\% of Russian oil exports implies a Russian
profit loss of 59 million USD per day in the short run (within a first
year) and 100 million USD per day in the long run (beyond a year).
This is equivalent to 1.2 and 2\% of Russia's (pre-war) GDP respectively;
and 32\% and 56\% of their (pre-war) military spending. 

An imposed price discount implies even greater losses for Russia.
A 20\% discount of the global oil price yields a Russian loss of 152
million USD per day in both the short and long run. This is equivalent
to 3.1\% of Russias (pre-war) GDP and 85\% of their (pre-war) military
spending.

A price discount of 20\% seems, at this moment, to be feasible. It
would correspond to around 20 USD/b -- less than what has been discussed
at G7 meetings and less than the Urals discount that has been in effect
since the Russian invasion. Even greater discounts may be feasible
and yield greater losses for Russia.

Furthermore, a price discount implies net gains for oil importers.
This implies that consensus of its value as an economic weapon should
be rather broad among oil importers (see Table \ref{Table: Regions quantity}).
In particular, our results suggest (Table \ref{Table: Regions quantity})
that it will be harder to get broad acceptance of a quantity restriction
compared to a price discount. It is particularly India who may dispute
a quantity restriction (they lose relatively more), and in second
place China. Furthermore, it is particularly India (second place China)
who gain from a price discount. 

A coordinated price discount is essentially equivalent to a buyers’
cartel. Thus follow its benefits. But, just like any cartel, the stability
of such an agreement is not obvious. Any one buyer has a self interest
in buying a little bit more from Russia at a somewhat higher price.
Similarly, Russia may choose not to export at the discounted price,
in which case the price-discount sanction becomes a de facto supply
restriction. The indifference curves in Figure \ref{Fig: P vs Q Russia}
are indicative of Russia's best response to a price discount. This
is particularly important to consider for the sanctioning side in
relation to the actual implementation of a price discount. Russia
always has the option to restrict its own exports of oil. The suggested
implementation of the price discount is to limit Russia's access to
tanker shipments unless it sells at or below the required price \citep{Reuters_2022}.
More precisely, tankers will not be able to use western insurance
companies if they ship Russian oil above the price cap. A large share
of the tankers use such insurance companies. A response for Russia
may then be to not use such tankers. In practice, Russia then chooses
a quantity restriction. The extent of this quantity restriction is
dependent on how much of Russia's transport needs that can be covered
by the remaining tanker market and to what extent buyers accept sanctioned
oil and non-western insurance. It is beyond this paper to fully analyze
this, we refer to \citet{spiro_import} for analysis of the tanker
market. 

A quantity restriction is equivalent to an agreed Cournot cartel.
The downside of this construction is that it implies losses for the
oil importers. In the short run the relative losses are smaller than
Russia's albeit in the same order magnitude. In the long run the losses
to oil importers are substantially smaller. Thus, while it may be
hard to get broad consensus around \textit{setting up} a quantity
restriction (in particular India loses), it should not be as hard
to keep it in place in the long term. Nevertheless it may be difficult
to argue among buyers for such a construction given these losses.
However, what drives the losses is a rise in the price of oil. And
this high price will make it less tempting to deviate from the agreement.
Unless, of course, Russia offers its oil at their own discount, in
which case the supply restriction is partly transformed into a de-facto
price discount. 

\pagebreak{}


\bibliography{refs}{}

\begin{thebibliography}{}

\bibitem[Allen, 2008]{allen2008domestic}
Allen, S.~H. (2008).
\newblock The domestic political costs of economic sanctions.
\newblock {\em Journal of Conflict Resolution}, 52(6):916--944.

\bibitem[Bachmann et~al., 2022]{bachmann2022if}
Bachmann, R., Baqaee, D., Bayer, C., Kuhn, M., L{\"o}schel, A., Moll, B.,
  Peichl, A., Pittel, K., and Schularick, M. (2022).
\newblock What if germany is cut off from russian energy?
\newblock {\em VoxEU. org}, 25.

\bibitem[Berger et~al., 2022]{berger2022potential}
Berger, E.~M., Bialek, S., Garnadt, N., Grimm, V., Salzmann, L., Schnitzer, M.,
  Truger, A., Wieland, V., et~al. (2022).
\newblock A potential sudden stop of energy imports from russia: Effects on
  energy security and economic output in germany and the eu.
\newblock Technical report, IMFS Working Paper Series.

\bibitem[BP, 2022]{bp_bp_2022}
BP (2022).
\newblock {BP} {Statistical} {Review} of {World} {Energy} 2022.
\newblock Technical report.

\bibitem[Chen et~al., 2019]{chen2019international}
Chen, Y.~E., Fu, Q., Zhao, X., Yuan, X., and Chang, C.-P. (2019).
\newblock International sanctions’ impact on energy efficiency in target
  states.
\newblock {\em Economic Modelling}, 82:21--34.

\bibitem[EC, 2022]{EC_2022}
EC (2022).
\newblock Eu restrictive measures against russia over ukraine (since 2014).

\bibitem[Erickson and Lazarus, 2014]{erickson2014impact}
Erickson, P. and Lazarus, M. (2014).
\newblock Impact of the keystone xl pipeline on global oil markets and
  greenhouse gas emissions.
\newblock {\em Nature Climate Change}, 4(9):778--781.

\bibitem[F{\ae}hn et~al., 2017]{faehn2017climate}
F{\ae}hn, T., Hagem, C., Lindholt, L., M{\ae}land, S., and Rosendahl, K.~E.
  (2017).
\newblock Climate policies in a fossil fuel producing country--demand versus
  supply side policies.
\newblock {\em The Energy Journal}, 38(1).

\bibitem[Fischhendler et~al., 2017]{fischhendler2017political}
Fischhendler, I., Herman, L., and Maoz, N. (2017).
\newblock The political economy of energy sanctions: insights from a global
  outlook 1938--2017.
\newblock {\em Energy Research \& Social Science}, 34:62--71.

\bibitem[Gardner and Psaledakis, 2022]{Reuters_2022}
Gardner, T. and Psaledakis, D. (2022).
\newblock U.s says g7 should soon unveil price cap level on russian oil, adjust
  regularly.

\bibitem[Gars et~al., 2022]{gars2022effect}
Gars, J., Spiro, D., and Wachtmeister, H. (2022).
\newblock The effect of european fuel-tax cuts on the oil income of russia.
\newblock {\em Nature Energy}, 7(10):989--997.

\bibitem[Gharehgozli, 2017]{gharehgozli2017estimation}
Gharehgozli, O. (2017).
\newblock An estimation of the economic cost of recent sanctions on iran using
  the synthetic control method.
\newblock {\em Economics Letters}, 157:141--144.

\bibitem[Hosoi and Johnson, 2022]{hosoi2022implement}
Hosoi, A. and Johnson, S. (2022).
\newblock How to implement an eu embargo on russian oil.
\newblock {\em CEPR Policy Insight}, 116.

\bibitem[Johnson, 1950]{johnson1950optimum}
Johnson, H.~G. (1950).
\newblock Optimum welfare and maximum revenue tariffs.
\newblock {\em The Review of Economic Studies}, 19(1):28--35.

\bibitem[Kennedy, 2022]{kennedy_2022}
Kennedy, C. (2022).
\newblock Putin's looming tanker crisis.

\bibitem[Lafrogne-Joussier et~al., 2022]{lafrogne2022beyond}
Lafrogne-Joussier, R., Levchenko, A., Martin, J., and Mejean, I. (2022).
\newblock Beyond macro: Firm-level effects of cutting off russian energy.
\newblock {\em Global Economic Consequences of the War in Ukraine Sanctions,
  Supply Chains and Sustainability}, page~8.

\bibitem[Rystad, 2022a]{rystad_energy_specialreport_Russia_2022}
Rystad (2022a).
\newblock {Oil Market Special Report:Russia}.

\bibitem[Rystad, 2022b]{rystad_energy_ucube_2022}
Rystad (2022b).
\newblock {UCube}.

\bibitem[Shapovalova et~al., 2020]{shapovalova2020russian}
Shapovalova, D., Galimullin, E., and Grushevenko, E. (2020).
\newblock Russian arctic offshore petroleum governance: The effects of western
  sanctions and outlook for northern development.
\newblock {\em Energy Policy}, 146:111753.

\bibitem[Spiro et~al., 2022]{spiro_import}
Spiro, D., Wachtmeister, H., and Gars, J. (2022).
\newblock Will russian income fall if the eu bans imports of russian oil?

\bibitem[Sturm, 2022]{sturm2022simple}
Sturm, J. (2022).
\newblock The simple economics of trade sanctions on russia: A policymaker’s
  guide.
\newblock Technical report, Working Paper, 9.4.

\bibitem[Treasury, 2022]{treasury_2022}
Treasury, U. D.~O. (2022).
\newblock Ukraine-/russia-related sanctions.

\bibitem[Weizent, 2022]{weizent_2022}
Weizent (2022).
\newblock Oil prices jump after eu leaders agree to ban most russian crude
  imports.

\bibitem[{World Bank}, 2022]{world_bank_exports_2022}
{World Bank} (2022).
\newblock Exports of goods and services (current {US}\$) - {Russian}
  {Federation} {\textbar} {Data}.

\end{thebibliography}
\bibliographystyle{apalike}

\end{document}